\documentclass[11pt,twocolumn]{article} 
\usepackage{epsf}
\usepackage{pazh_eng}
\usepackage[english]{babel}
\usepackage{graphicx}
\usepackage[round,sectionbib]{natbib}
\usepackage{amssymb}
\usepackage[small]{caption2}
\tightenlines
\usepackage[left=2cm,right=2cm,top=2cm,bottom=3cm,bindingoffset=0cm]{geometry}

\def\SSobj{SS\,433}

\def\asprv{ASPRV}
\begin{document}

\twocolumn[
\begin{@twocolumnfalse}
\title{\bf X-ray Variability of SS 433: Evidence for Supercritical Accretion}
	
\author{\bf K.\,E.\,Atapin\affilmark{1,2*}, S.\,N.\,Fabrika\affilmark{2,3}}
	
\affil{
	{\it $^1$Sternberg Astronomical Institute, Lomonosov Moscow State University,\\ Universitetskii pr. 13, Moscow, 119991 Russia}\\ 
	{\it $^2$Special Astrophysical Observatory, Russian Academy of Sciences,\\ Nizhnii Arkhyz, Karachai-Cherkessian Republic, 369167 Russia}\\
	{\it $^3$Kazan Federal University, ul. Kremlevskaya 18, Kazan, 420000 Russia}}
	
	\vspace{-4mm}
	\center{Received January 12, 2016}

\begin{abstract}
\noindent We study the X-ray variability of \SSobj\ based on data from the ASCA observatory and the MAXI and RXTE/ASM monitoring missions. Based on the ASCA data, we have constructed the power spectrum of \SSobj\ in the frequency range from $10^{-6}$ to 0.1\,Hz, which confirms the presence of a flat portion (flat-topped noise) in the spectrum at frequencies $3\times 10^{-5} - 10^{-3}$\,Hz. The periodic variability (precession, nutation, eclipses) begins to dominate significantly over the stochastic variability at lower frequencies, which does not allow the stochastic variability to be studied reliably. The best agreement with the observations is reached by the model with the flat portion extending to $9.5\times10^{-6}$~Hz and a power-law spectrum with index of 2.6 below that frequency. The jet nutation with a period of about three days suggests that the time for the passage of material through the disk is less than this value. Therefore, at frequencies below $4\times10^{-6}$~Hz, the power spectrum probably does not reflect the disk structure. It may depend on other factors, for example, a variable mass accretion rate supplied by the donor. The flat portion may arise from a rapid decrease in the viscous time in the supercritical or radiative disk zones. It could be related to the variability of the X-ray jets that are formed in the supercritical region.\\
\end{abstract}

\vspace{-0.2cm}\hspace{0.68cm}
\vspace{2.0cm}
\end{@twocolumnfalse}]

\section*{INTRODUCTION}
\noindent \SSobj\ is the only known super-accretor in the Galaxy. This is a binary system; it consists of a black hole with a mass of $\sim10M_\odot$ and a massive Roche lobe-filling star \citep[for a review, see][]{Fabrika2004}. The accretion rate is estimated to be $M_0\sim300 M_{Edd}$, where $M_{Edd}$ is the Eddington accretion rate. A supercritical accretion disk, whose key feature is the presence of a strong wind, is formed in the system \citep{ShakSun1973,Fabrika1997,Brinkmann2005}. This wind arising in the innermost parts of the
\noindent\rule{8cm}{1pt}\\
{\mbox{\hspace*{0.5cm}$^*$E-mail\hspace*{0.5cm} \texttt{atapin.kirill@gmail.com}}}\\ 
disk, below the spherization radius \citep{ShakSun1973}, forms an optically thick wind funnel in the shape of a hollow cone \citep{AtapinSS433var2015}. Relativistic jets moving with a speed of $0.26c$ are formed along the funnel axis.

The bolometric luminosity of \SSobj\ is $\sim10^{40}$ erg~s$^{-1}$ \citep{Cherepashchuk2002}. Almost all of the energy in the system is released in the supercritical accretion disk, initially in the X-ray range. This is confirmed by the colossal kinetic energy of the jets \citep[$\sim10^{39}$ erg~s$^{-1}$,][]{Panferov1997}. However, the optically thick wind covers the hard emission source in the inner parts of the disk. Therefore, the already thermalized radiation with its maximum in the ultraviolet reaches the observer \citep{Dolan1997}. The observed X-ray emission ($\sim10^{36}$ erg~s$^{-1}$) comes mainly from the cooling jets \citep{MedvFabr2010}.

The spectrum of the X-ray jets has been well studied \citep{Brinkmann2005,Filippova2006,Marshall2013,Khabibullin2016}; it is formed by the adiabatic and radiative cooling of the jet gas. In addition to the jets prevailing in the range 2--5~keV, the X-ray spectrum of \SSobj\ also exhibits the soft ($\leq 1$~keV) and hard ($\geq 8$~keV) components associated with the wind \citep{MedvFabr2010}. The soft component is the intrinsic thermal radiation from the wind funnel wall; the hard component is the radiation emerging from the inner accretion disk and reflected toward the observer by the outer parts of the funnel. Evidence that the optical and ultraviolet emissions also originate in the outer parts of the funnel via the fluorescence of X-ray emission coming from its inner parts has been found \citep{Revn2004,AtapinSS433var2015}.

The jets and the funnel precess with a period $P_{pr}\approx 162$ days. At the precession phase $\psi=0$, the accretion disk is maximally turned toward the observer. At this time, \SSobj\ appears brightest and hottest in the optical range; its temperature is 50\,000--70\,000~K \citep{Dolan1997}. At phases $\psi\approx0.34$ and 0.66, the accretion disk is seen edge-on, and the funnel is closed for the observer. \SSobj\ is an eclipsing binary. Its orbital period is $P_{orb}\approx13.08$ days. The center of the accretion disk eclipse by the donor star corresponds to the orbital phase $\varphi=0$.

Since the plane of the accretion disk does not coincide with the orbital plane of the star, the tidal effect of the donor star on the disk leads to nutation. The nutation has a period $P_{nut}\approx6.29$ days \citep{Goranskij1998} and is clearly seen in both optical and X-ray ranges. The gravitational field of the star changes the inclination of the outer parts of the disk to the line of sight, which manifests itself in the optical range as a change in the system's magnitude by 0\fm2 \citep{Goranskij2011}. Spectroscopic observations show that the jets at these times deviate by an angle up to $\approx3^\circ$ \citep{Fabrika2004}. The clear relationship between the tidal effect on the disk, mainly on its peripheral regions, and the nodding of the jets forming in its innermost parts suggests that the disk must be restructured in a time $\lesssim P_{nut}$ \citep{Revn2006}.Thus, the time for the passage of material through the disk must be less than three days \citep{Fabrika2004}.

The disk structure can be judged by studying the stochastic variability of the system. Based on optical, radio, and X-ray observations, \cite{Revn2006} constructed the power spectrum in the frequency range $10^{-8}-10^{-2}$~Hz and found that, on average, it was satisfactorily described by a single power law $P\propto \nu^{-\beta}$ with $\beta\approx1.5$. This pattern of variability was interpreted as noise arising from the fluctuations in viscosity $\alpha$ at various radii of the accretion disk \citep{Lyubarskii1997}. The authors also suggested that on time scales $\sim l/c$, where $l$ is the funnel size, the power spectrum must exhibit a break associated with the smearing of the variability in the funnel. This break was detected by \cite{Burenin2011} in the optical range near $10^{-3}$~Hz.

Analyzing the RXTE observations of \SSobj, we detected this break in the X-ray range \citep{AtapinSS433var2015}. In addition, we found the shape of the power spectrum and the presence of a break to depend on the precession phase: the break appears when the funnel is turned toward the observer. If the disk is seen edge-on, then there is no break and the entire power spectrum has an index $\beta\approx 1.34$. At frequencies below the break, the X-ray power spectrum is flat, i.\,e., its amplitude does not depend on the frequency \citep[on the whole, the optical power spectrum is steeper and has an index of $\sim 1$ at frequencies below the break;][]{Burenin2011}. Since this flat portion (flat-topped noise) is observed at the precession phase when the deepest parts of the funnel are visible, it was concluded that it is determined by some noise process in the innermost parts of the supercritical accretion disk. We assumed that this noise was generated at the spherization radius and was related to a sharp decrease in the time for the passage of material through the disk in its supercritical part. 

The duration of the RXTE observations did not allow us to trace the shape of the power spectrum of \SSobj\ below $10^{-4}$~Hz. It was assumed that at lower frequencies the flat portion would end and the power spectrum would then again become a power-law one \citep{Revn2006}. In this paper, we use the ASCA X-ray observations, which allows us to extend the power spectrum to $10^{-6}$~Hz and to impose new constraints on the extent of the flat portion. We also use the monitoring observations of \SSobj\ to study its power spectrum at lower frequencies.

\vspace*{1cm}
\section*{OBSERVATIONS}
\noindent The objective of this paper was to study the power spectra at frequencies below $10^{-4}$~Hz. Long homogeneous observations with a duration of at least one day are needed to cover this frequency range. There are such uniquely long observations in the ASCA archive. In addition to the ASCA observations, we also analyzed the MAXI and RXTE/ASM monitoring data.

\subsection*{ASCA Data}
\noindent The ASCA (Advanced Satellite for Cosmology and Astrophysics) observatory was in orbit from February 20, 1993, to March 2, 2001. The observatory was equipped with four X-ray telescopes with detectors of two types: two gas imaging spectrometers (GIS) and two solid-state imaging spectrometers (SIS) \citep{TanakaASCA1994}. The operating ranges for both types of detectors are approximately identical: 0.7--10~keV for GIS and 0.4--10~keV for SIS. 

There are a total of 31 observations for \SSobj\ in the ASCA archive, two of which have a duration considerably longer than one day. The longest observation (ObsID 48002000) lasted 11.7 days from March 23 to April 3, 2000. Over this period, the precession phase of the system\footnote{We used the ephemerides from \cite{Goranskij2011} to calculate the orbital and precessional phases.} changed from $\psi\approx0.04$ to 0.11; thus, at the beginning of the observations, the accretion disk of \SSobj\ was turned toward the observer almost by the maximum angle. An eclipse of the X-ray source by the donor star (the eclipse center occurred on March 29) that lasted about three days occurred approximately in the middle of the observation. The second shorter observation (ObsID 44011000) has a duration of three days. However, it completely fell within the eclipse, and we did not use it.

\begin{figure*}
\center
\includegraphics[width=0.6\textwidth]{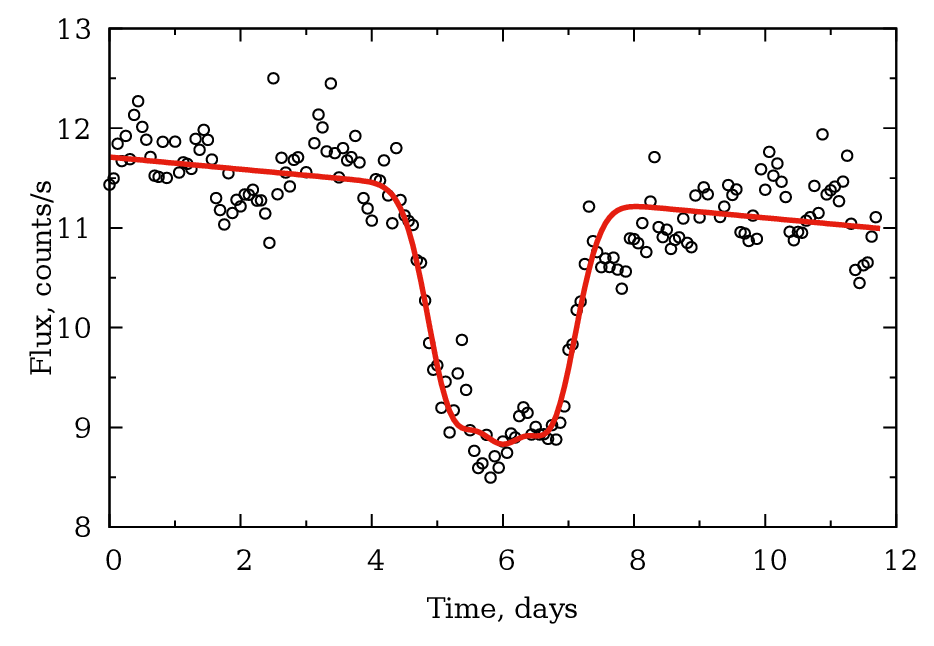}
\caption{Light curve of the twelve-day ASCA observation of \SSobj\ binned in 5400~s. The energy range is 0.7--10~keV. The flux errors are comparable to the circle sizes. The accretion disk eclipse by the donor star is responsible for the dip at the center of the light curve. The solid line is the fit to the eclipse and precession trend by a smooth function.}
\label{fig:asca_lc}
\end{figure*}

In this paper, we analyze the preprocessed light curves of the twelve-day observation from the archive at the NASA site\footnote{http://heasarc.gsfc.nasa.gov/FTP/}. We used only the data from the more reliable GIS instrument. To increase the signalto-noise ratio, the light curves from both GIS were added. The mean flux of the resulting light curve was about 11.4 photons/s in the 0.7--10~keV, the error was 3.5 photons/s, and the temporal resolution was 1~s. During the eclipse, the X-ray flux drops by 20\% (Fig.~\ref{fig:asca_lc}). A gradual decrease in the flux associated with a change in the inclination of the accretion disk funnel due to precession is also observed in the light curve.

Since the ASCA observatory was in a low orbit with a period of about 90~min, the light curves that we analyzed are not continuous. The light curves have periodic gaps due to the occultation of the observed object by the Earth and the passage of the satellite through the South Atlantic Anomaly. The lengths of the continuous light-curve segments do not exceed 3000~s; the total fraction of gaps is about 50\%. In the subsequent analysis, we pay attention to the gaps and their influence on the shape of the power spectra.

\subsection*{RXTE Data}
\noindent The RXTE (Rossi X-ray Timing Explorer) observatory was launched on December 30, 1995, and had operated until January 5, 2012. It was equipped with three instruments: the proportional counter array (PCA), the HEXTE (High Energy X-ray Timing Experiment) scintillation detector, and the all-sky monitor (ASM). The PCA was the main instrument of the observatory. It operated in the energy range 2--60~keV and had a high sensitivity and a large collecting area ($\sim6500$~cm$^2$). The HEXTE detector had a lower sensitivity, but it was designed for higher energies ($15-250$~keV). These two instruments were used for the pointing observations and operated simultaneously.

The ASM operated independently from other instruments and was designed to monitor the sky in the energy range 1.5--12~keV divided into three channels: 1.5--3, 3--5, and 5--12~keV. The instrument included three wide-field shadow cameras that surveyed up to 80\% of the sky per one 1.5~h rotation, providing regular observations of more than 300 X-ray sources. The total collecting area of the cameras is 90~cm$^2$. When scanning, the camera exposed each source for 90~s. The exposures of a specific source could either follow one another or be separated by a considerable time interval.

We analyzed the preprocessed light curves of \SSobj\ publicly accessible at the ASM site\footnote{http://xte.mit.edu/asmlc/ASM.html}. The light curves of two types are accessible at the site: the original ones, in which each point corresponds to one individual 90-s exposure, and the ones binned into one-day bins. The original light curves have no uniform time step: the start times of the exposures in them are scattered randomly. For the convenience of comparing the results with the MAXI data (see below), we analyzed the original light curve rebinned on a grid with a constant step of 5400~s; if several exposures fell within a given time bin, then we averaged the flux; if none fell within this bin, then we marked this point as a gap. As a result, we obtained a light curve with a constant time step containing random gaps. The total fraction of gaps is about 65\%. The duration of the ASM observations is more than 15 years, from MJD\,50088 to MJD\,55846. We used the data in the total range 1.5--12 keV. The mean flux was 0.3 photons/s; the error was 1.4 photons/s. 

After the completion of the RXTE mission, the mission-long data were published\footnote{http://heasarc.nasa.gov/docs/xte/recipes/mllc\_start.html}, in which the estimated mean (over the time of this PCA observation) flux from the object under study is assigned to each PCA observation. Because of the high PCA sensitivity, these data have a very high accuracy but contain few points. There are a total of 130 such measurements over 15 years of observations for \SSobj. Nevertheless, we also used these data by combining them into a light curve with a uniform step of 1 day. The derived light curve has a very large percentage of gaps, more than 97\%. The 2--20~keV flux changes from $\sim15$ to $\sim30$ photons/s, depending on the system's precession phase. The accuracy is 0.5\% or better.

\subsection*{MAXI Data}
\noindent MAXI (Monitor of All-sky X-ray Image) is a monitoring mission designed as a replacement of ASM/RXTE. The instrument is part of the Japanese experimental module onboard the International Space Station \citep{MatsuokaMAXI2009}; it has been operating since August 2009. Just as ASM, MAXI surveys the sky every 1.5~h. The total effective collecting area of the detector is about 5000~cm$^2$; the energy range is 2--20~keV. 

The up-to-date light curves are periodically supplemented with new data and are published on the home page of the project\footnote{http://maxi.riken.jp/top/}. The version of the light curve we use covers the range of Julian dates from MJD\,55058 to MJD\,57052. The time step is fixed, 5400~s, but there are random gaps. The total fraction of gaps is about 30\%. The mean 2--20~keV flux was 0.03 photon/s; the error was 0.06 photon/s.

\vspace*{1cm}
\section*{RESULTS}
\subsection*{Power Spectra from ASCA Data}
\begin{figure*}
\center
\includegraphics[width=0.50\textwidth]{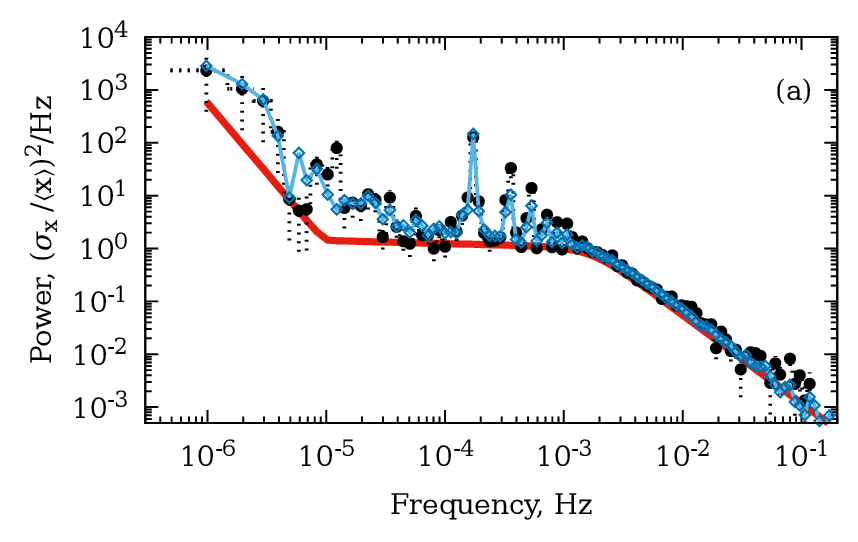}
\hspace{-0.3cm}
\includegraphics[width=0.50\textwidth]{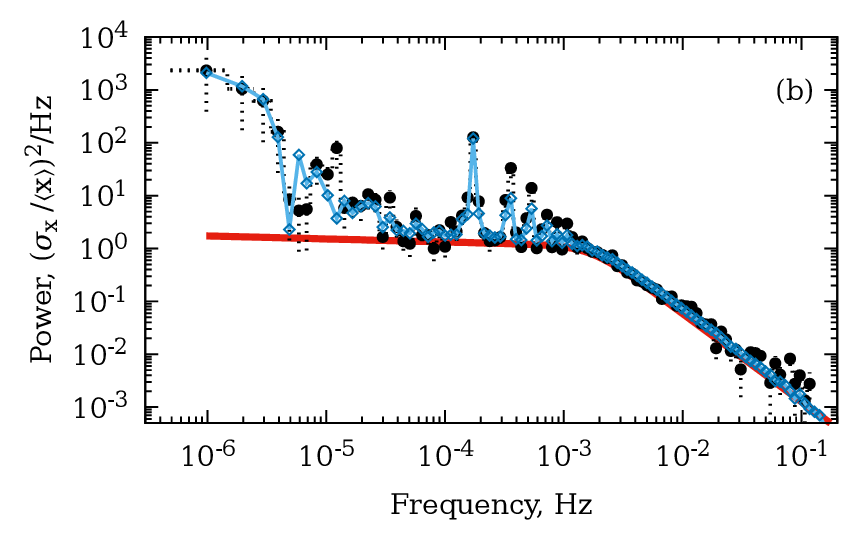}
\caption{The black circles and dotted line indicate the observed power spectrum of \SSobj\ in the fractional rms squared normalization in the range 0.7--10~keV. The peaks in the frequency range $10^{-4}-10^{-3}$~Hz are artifacts and are caused by gaps in the observation. The thick red (dark-gray) line indicates the initial model of intrinsic accretion disk variability (see the text). The synthetic Monte Carlo model obtained by adding the eclipse-precession trend to the initial model and the distortions associated with the gaps in the observations is indicated by the blue (light-gray) line and diamonds. (a) The model with a break at a frequency of $9.5\times 10^{-6}$~Hz, at which the best agreement with the observations is achieved. (b) The model for the case where the flat 	portion extends to the lowest frequencies. Despite the significant difference between the initial models, the synthetic models turned out to be almost identical, which makes the search for the position of the low-frequency break difficult.}
\label{fig:asca_pds}
\end{figure*}

\noindent All of the observational data with which we had to work (ASCA and the light curves from the monitors) have a significant fraction of gaps. It is well known that the gaps in the observations, especially the periodic ones, as, for example, in the ASCA data, can distort the shape of the power spectrum due to the so-called red noise leakage \citep{Priestley}. This effect is that strong side lobes into which the power is redistributed appear in the response function at frequencies corresponding to the period of gaps. As a result of this redistribution, spurious peaks that do not reflect the real variability of the object under study can appear in the power spectrum. The more complex the structure of gaps, the more complex the pattern of redistribution.

Nevertheless, we do not attempt to remove these distortions from the observations and do not use the methods often applied in analyzing unevenly spaced time series \citep{Lomb1976,Scargle1982}. Instead we apply the Monte Carlo method to introduce the same distortions as those in the observed power spectrum into the model to subsequently compare the distorted data with the distorted model. A similar Monte Carlo method was proposed by \cite{Uttley2002} to analyze the power spectra of active galactic nuclei \cite[the PSRESP method, see also][]{Done1992,GreenMcHardyDone1999}.

To reduce the computation time, we used a simple scheme to construct the power spectra. From the light curve we subtracted the mean flux and replaced the gaps by zeros, which allowed us to use the fast Fourier transform algorithm. The power spectrum was then normalized in the fractional rms squared normalization \citep{vanderKlis1997,Vaughan2003} and rebinned with a constant step on a logarithmic scale. No additional corrections were applied. 

The observed ASCA power spectrum of \SSobj\ is shown in Fig.~\ref{fig:asca_pds}. The peaks at $1.8\times10^{-4}$~Hz and multiples of this frequency correspond to the orbital period of the observatory and are artifacts. These peaks are caused by the power leakage and do not represent the real variability of the object. Nevertheless, despite the distortions, it can be seen from the figure that there is actually a flat portion in the power spectrum of \SSobj: the ASCA observations confirm the result that we obtained previously from RXTE data \citep{AtapinSS433var2015}. The flat portion in Fig.~\ref{fig:asca_pds} extends from $\sim 10^{-3}$~Hz to at least $3\times10^{-5}$~Hz.

A rise is observed in the power spectrum at lower frequencies. However, the variability introduced by the eclipse and precession begins to dominate over the intrinsic accretion disk variability at these frequencies. It can be seen from Fig.~\ref{fig:asca_lc} that the X-ray flux drops by 20\% during the eclipse. The precessional variations at a given phase (just after the precessional maximum) manifest themselves in the light curve as a quasi-linear descending trend (Figs.~\ref{fig:asca_lc} and. \ref{fig:ph_prec}). Both these effects are geometric in nature and are not associated with the accretion disk variability. Nevertheless, they make a major contribution to the variability on a time scale of one day or more in the observed light curve. 

The Monte Carlo method that we used to model the true shape of the power spectrum of the accretion disk by taking into account the influence of gaps and the contribution of the eclipse and precession includes the following steps. First, the \textit{initial model} $M_\nu(\theta_i)$ of the power spectrum described by a set of parameters $\theta_i$ is specified. As the initial model we used a function consisting of three power-law segments with two breaks (Fig.~\ref{fig:asca_pds}a, the thick solid line):
\begin{equation}
M_\nu \propto \frac{ \sqrt{1+(\nu_1/\nu)^{2\beta_1}} }{\nu^{\beta_2} \sqrt{1+(\nu/\nu_2)^{2\beta_3}}},
\label{eq:pdsdbl}
\end{equation}
where $\nu$ is an independent variable (frequency), while $\nu_1$, $\nu_2$, $\beta_1$, $\beta_2$, and $\beta_3$ are the model parameters. The parameters $\nu_1$ and $\nu_2$ describe the positions of the low- and high-frequency breaks, where a ``switch'' between the power-law segments occurs. The three other parameters $\beta_1$, $\beta_2$, and $\beta_3$ define the power-law indices of the segments. The index is $\beta_2$ at medium
frequencies in the interval between the breaks; the index
asymptotically tends to $\beta_1+\beta_2$ at low frequencies
$\nu \ll \nu_1$ and to $\beta_2+\beta_3$ at high ones $\nu \gg \nu_2$.

Next, the synthetic light curves corresponding to the initial model are generated. Since the power spectrum contains information only about the amplitude of harmonics, but there is no information about the phase, by specifying the phase by a random number generator, we can obtain an infinite number of various synthetic light curves each of which will have a power spectrum of the same specified shape. A generation algorithm convenient for application in practice, which uses the inverse Fourier transform, was proposed by \cite{Timmer1995}. The synthetic light curves must have the same temporal resolution and mean flux as the real light curve. Then, gaps are introduced into each synthetic light curve at the same places where they were in the observed light curve, and the eclipse and precession trend are added. The power spectra are computed, with the average level being subtracted from each light curve and the gaps being filled with zeros. Subsequently, these individual power spectra are normalized and binned in the same way as the observed one, and a synthetic power spectrum (synthetic model) is obtained after their averaging. When averaging a large number of individual power spectra, the random noise weakens, while the systematic distortions due to the gaps remain unchanged.

\begin{figure}
	\center
	\includegraphics[width=0.48\textwidth]{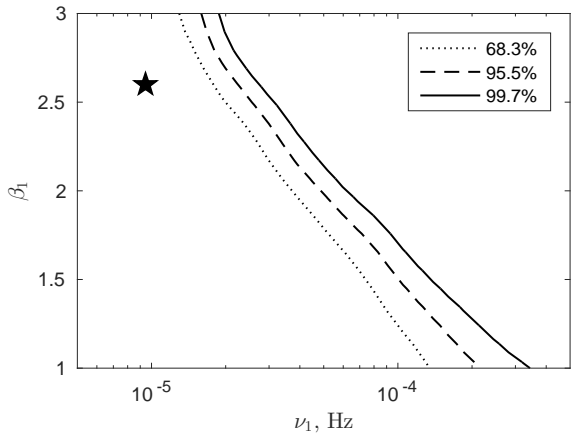}
	\caption{The contours corresponding to the 68.3\%, 95.5\%, and 99.7\% confidence intervals of the parameters $\nu_1$ (the position of the low-frequency break) and $\beta_1$ (the slope of the power spectrum at the lowest frequencies). The star marks the parameters of the model that provided the best agreement with the observations (Fig.~\ref{fig:asca_pds}a). It follows from the figure that we failed to reveal any constraints on the parameters $\nu_1$ and $\beta_1$.}
	\label{fig:asca_mlfunc}
\end{figure}

\begin{figure*}
	\center
	\includegraphics[width=0.68\textwidth]{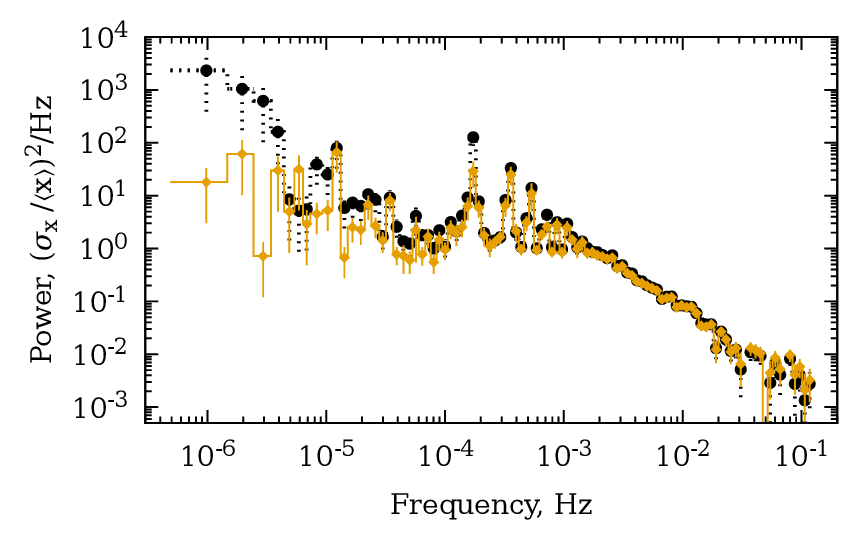}
	\caption{The black circles and dotted line indicate the observed power spectrum (the same as that in Fig.~\ref{fig:asca_pds}). The yellow (gray) line indicate the power spectrum obtained by correcting the observational data themselves for the eclipse-precession trend.}
	\label{fig:asca_pds_ecl}
\end{figure*}

The synthetic model obtained in this way from the initial model takes into account the distortions inherent in the observations. To estimate the extent to which the synthetic model $P_k(\theta_i)$ corresponds to the observed power spectrum $I_k$, we calculated the maximum likelihood statistic:
\begin{equation}
S(\theta_i)=2\sum\limits_{k=1}^{N}\left(\ln[P_k(\theta_i)] + {I_k\over P_k(\theta_i)}\right),
\label{eq:pdsstat}
\end{equation}
which is suited for the approximation of single power spectra better than the least-squares method \citep{Stella1994,Vaughan2005}. The functional $S$ should be minimized by varying the parameters and normalization of the initial model $M_\nu(\theta_i)$. In this case, for each new value of the parameters, we have to repeat all of the actions needed to compute the synthetic power spectrum. When the minimum $S_0$ of the functional will be found, and the synthetic power spectrum will achieve the best agreement with the observed one, we may then assert that the initial model is a good approximation of the true power spectrum (the intrinsic disk variability). 

The parameters $\nu_1$ and $\beta_1$ describing the low frequency break are of greatest interest to us. We previously measured the position of the high-frequency break and the power-law slope above and below this break \citep{AtapinSS433var2015} based on RXTE: $\nu_2 = 1.7\times10^{-3}$~Hz, $\beta_2 = 0.06$, and $\beta_3 = 1.61$. In addition, since the high-frequency part of the power spectrum is less affected by distortions, its parameters can be reliably measured from the available data without resorting to complex modeling.

To measure the high-frequency part of the power spectrum based on ASCA data, we did the following. From the twelve-day light curve we excluded the central three days when the eclipse occurred (Fig.~\ref{fig:asca_lc}) and divided the remaining part into segments about 3~ks in length corresponding to the times between the occultations of the object by the Earth during which the light curve was continuous. We obtained a total of $\sim100$~bins. For each bin, we calculated the individual power spectra, which were then averaged. Since there are virtually no gaps within these segments, the resulting power spectrum has no distortions, but it contains no frequencies below $3\times10^{-4}$~Hz. 

Due to the averaging over a large number of bins, the distribution of errors in the final power spectrum is very close to the Gaussian one, and the standard least-squares method can be applied for its approximation \citep{Papadakis1993}. As a result of the approximation, we obtained $\beta_3= 1.62\pm0.09$ and $\nu_2 = (1.54\pm0.32) \times 10^{-3}$~Hz, with the slope of the (quasi) flat portion having been fixed at $\beta_2 = 0.06$ that we obtained previously from RXTE data \citep{AtapinSS433var2015}. Good agreement of the parameters derived from the ASCA data with the RXTE observations made mostly in 2004 suggests a high stability of the power spectra for \SSobj. Hereinafter, when modeling the power spectrum in the entire frequency range, we assumed these three parameters ($\beta_2$, $\beta_3$, and $\nu_2$) describing the high frequency break to be known, constant and did not vary them.

Thus, to determine the parameters of the low frequency break in the power spectrum of \SSobj, we used the model (\ref{eq:pdsdbl}) in which two parameters were free, while the remaining parameters were fixed at their measured values. For each pair of free parameters $\nu_1$ and $\beta_1$, we generated $\sim 1000$ synthetic light curves using the algorithm by \cite{Timmer1995}. To avoid the additional distortions due to the application of the inverse Fourier transform, we generated synthetic light curves twice longer than real light curve \citep{Uttley2002,MiddletonDonePDSsym2010}. Subsequently, these superfluous parts were cut off. We then multiplied each light curve by the eclipse-precession trend that we obtained by fitting the observed light curve by a smooth function (Fig.~\ref{fig:asca_lc}) and introduced the pattern of gaps. Next, we computed the synthetic model and the maximum likelihood statistic $S$. 

Figure~\ref{fig:asca_pds}a shows the best-fit model ($\nu_1 \approx 9.5 \times 10^{-6}$~Hz and $\beta_1\approx2.6$), but the modeling result depends very weakly on the parameters of the initial model. The systematic variability introduced by the eclipse and precession prevails at low frequencies and completely dominates over the intrinsic stochastic variability of the accretion disk. Figure~\ref{fig:asca_pds}b shows the model where there is no low-frequency break and the flat portion extends to the lowest frequencies. Despite the significant difference between the initial models in Figs.~\ref{fig:asca_pds}a and \ref{fig:asca_pds}b, the synthetic power spectra taking into account the contributions of the trends and the influence of the gaps turned out to be nearly identical: the position of the low-frequency break cannot
be reliably determined from the observations.

\begin{figure*}
	\center
	\includegraphics[width=0.32\textwidth]{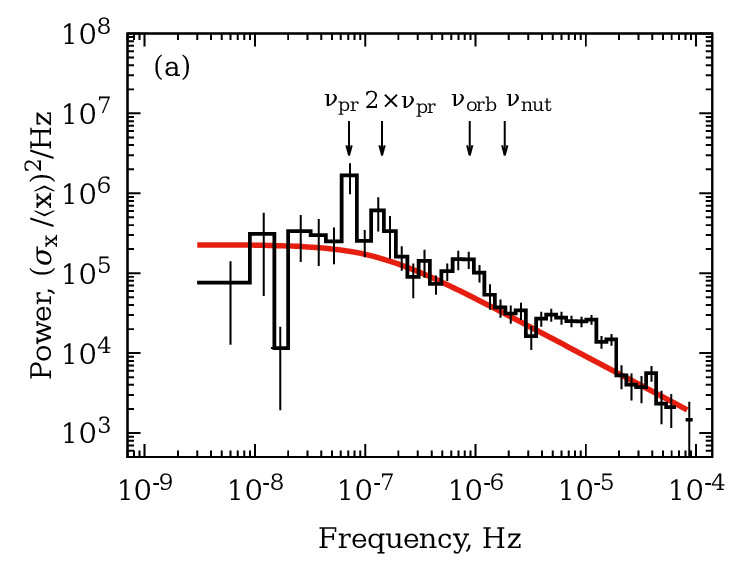}
	\includegraphics[width=0.32\textwidth]{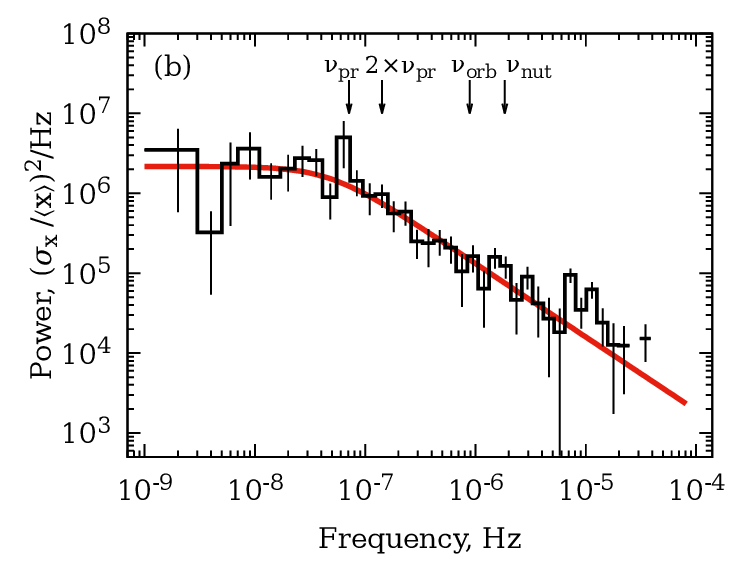}
	\includegraphics[width=0.32\textwidth]{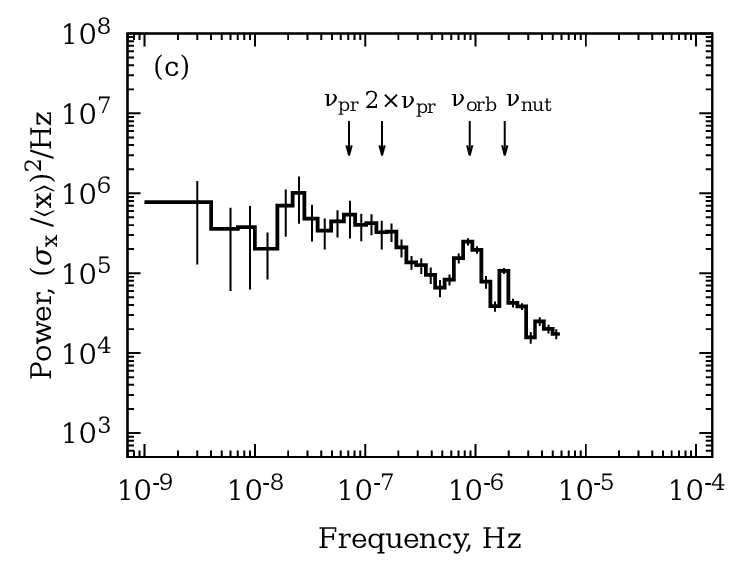}
	\caption{Power spectrum of \SSobj\ constructed from the MAXI (2--20~keV) (a) and RXTE/ASM (1.5--12~keV) (b) monitoring data as well as the RXTE/PCA mission-long data (2--60~keV) (c). The arrows mark the peaks associated with the three main periods of the system: precessional, orbital, and nutational. The solid line indicates the power-law fit to the continuum of the power spectrum.}
	\label{fig:pds_long_data}
\end{figure*}

It is well known \citep{Stella1994} that if a model has two free parameters, then the difference $\tilde{S}(\theta_i) = S(\theta_i)-S_0$, where $S_0$ is the minimum of the functional at which the best agreement of the data with the model is achieved, is a random variable obeying a chi-square distribution with two degrees of freedom. We calculated $\tilde{S}$ for break frequencies $\nu_1$ from $10^{-6}$ to $10^{-3}$~Hz and values of the slope $\beta_1$ from 0 to 3. To eliminate the influence of the high-frequency part of the power spectrum, we performed the summation in Eq.~(\ref{eq:pdsstat}) only over the frequencies below $10^{-4}$~Hz. Figure~\ref{fig:asca_mlfunc} shows the contours corresponding to various confidence intervals of the parameters. It can be seen from the figure that only the cases of either excessively steep slopes or excessively high break frequencies go beyond $3\sigma$ (99.7\%). At a break frequency $\nu_1 < 10^{-5}$~Hz, all slopes turn out to be equally admissible and do not go beyond $1\sigma$.

In the above modeling, we added the variability associated with the eclipse and precession to the model, while the original observational data remained unchanged. Figure~\ref{fig:asca_pds_ecl} shows another version of the power spectrum obtained by removing the eclipse-precession trend directly from the observed light curve itself. We see that an increase in the power is observed at frequencies below $\sim3\times10^{-5}$~Hz; nevertheless, the power here is appreciably lower than that in Fig.~\ref{fig:asca_pds}. This rise may reflect the real shape of the power spectrum and also be related to the trend correction. This can also be caused by real effects, for example, by a change in the amount of material supplied by the donor.

Thus, based on the ASCA data, we confirmed the result obtained previously from the RXTE observations \citep{AtapinSS433var2015}: at the phase of maximal opening of the accretion disk and the funnel to the observer, a flat portion appears in the power spectrum of \SSobj. It extends at least from $\sim3\times10^{-5}$ to $1.5\times 10^{-3}$~Hz. At lower frequencies, the power spectrum can both become a power law again and remain flat. No firm conclusion about the extent of the flat portion can be drawn from these data.

\subsection*{Power Spectra from Monitoring Data}

\noindent The power spectra constructed from the monitoring observations are shown in Fig.~\ref{fig:pds_long_data}. Since we reduced all light curves to a uniform time step, we applied the same method using the fast Fourier transform as that in our analysis of the ASCA data to compute the power spectra. To take into account the decrease in power (normalization) due to the presence of random gaps, we multiplied the power spectra by $N^2/(N-N_{gap})^2$, where $N_{gap}$ is the number of gaps \citep{SugimotoMAXI2014}.

The power spectrum constructed from the MAXI observations is shown in Fig.~\ref{fig:pds_long_data}a: the peaks at the precession frequency, twice the frequency, and the orbital frequency are clearly seen. The power spectrum has a power-law shape at frequencies above the precession peak; a bend is observed below the peak. Fitting by a power-law model with a break yielded the following parameters: the position of the break
$\nu_{br} \sim 2\times10^{-7}$~Hz and the slope $\beta = 0.87\pm 0.17$ at frequencies $10^{-7}-10^{-4}$~Hz. During our fitting, we assumed the power spectrum to be flat at frequencies below the break \citep{Revn2006}.

Figure~\ref{fig:pds_long_data}b shows the RXTE/ASM power spectrum. It clearly shows only the precession period; there are no peaks at other frequencies. Our fitting yielded the parameters $\nu_{br} \sim 4\times 10^{-8}$~Hz and $\beta = 0.92\pm 0.14$. The power-law slope at frequencies above the break in the ASM power spectrum coincided, within the error limits, with the MAXI results. However, the positions of the break differ: the break in the longer ASM observations is at a lower frequency. The amplitudes also differ. Since the relative variability amplitude of \SSobj\ is higher in the hard range \citep{AtapinSS433var2015}, one would expect the MAXI (2--20~keV) power spectrum to be higher than the ASM (1.5--12~keV) power spectrum. However, the opposite is true in Fig.~\ref{fig:pds_long_data}. This is probably because the ASM sensitivity is low; as a result, these data have large measurement errors. In addition, when the power spectrum is computed in fractional rms squared normalization \citep{vanderKlis1997}, a division by the square of the source mean flux takes place. If the flux after the background subtraction is underestimated, then the power spectrum in fractional rms squared normalization will be overestimated. Therefore, we assume the data from the more modern and sensitive MAXI instrument to be more reliable and to better reflect the shape of the power spectrum for \SSobj.

The power spectrum from the RXTE/PCA mission long data is shown in Fig.~\ref{fig:pds_long_data}c. Because of the small number of points in the light curve and the very large fraction of gaps, this power spectrum is affected by strong distortions, and we did not fit it. The precession period in these data turned out to be indistinct, but the peak at the orbital frequency and the peak at twice the frequency, probably related to nutation, are clearly seen.

\begin{figure}
	\center
	\includegraphics[width=0.45\textwidth]{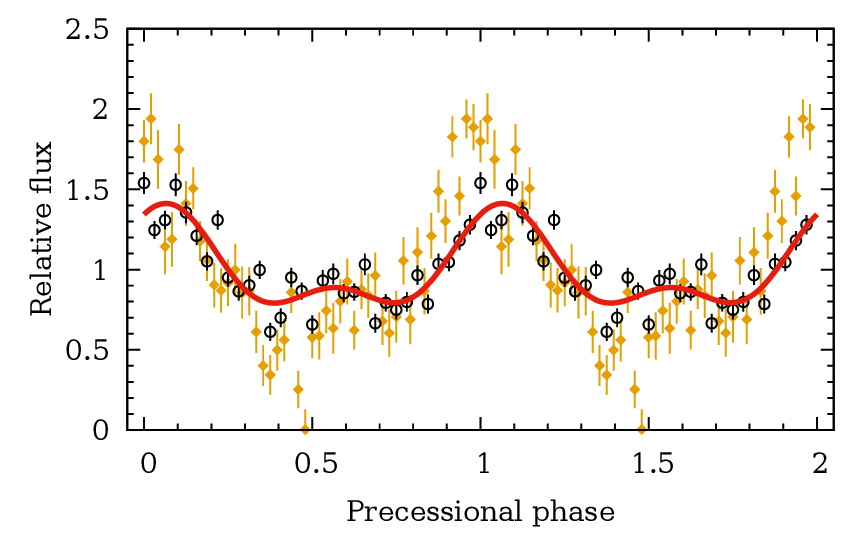}
	\caption{Precessional epoch folded light curves of \SSobj\ constructed from the MAXI (black circles) and ASM (yellow diamonds) observations. The solid line is the fit to the MAXI data by two Fourier harmonics.}
	\label{fig:ph_prec}
\end{figure}

\begin{figure*}
	\center
	\includegraphics[width=0.32\textwidth]{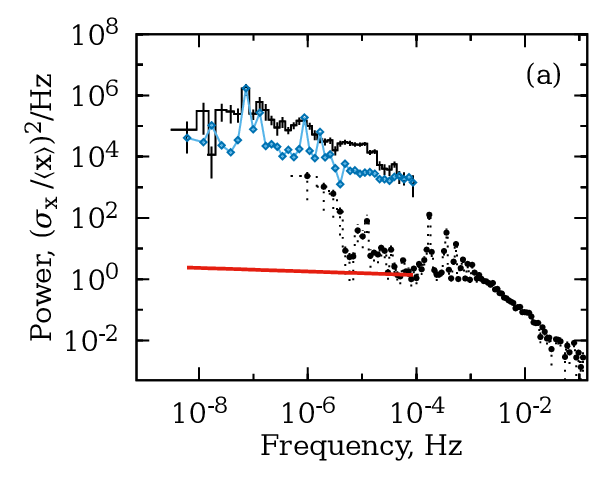}
	\includegraphics[width=0.32\textwidth]{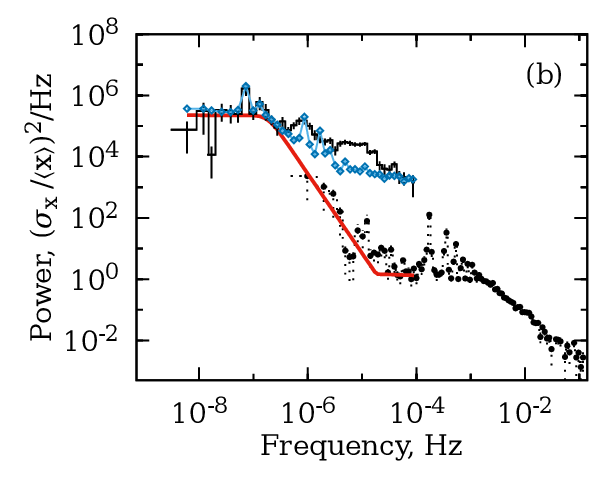}
	\includegraphics[width=0.32\textwidth]{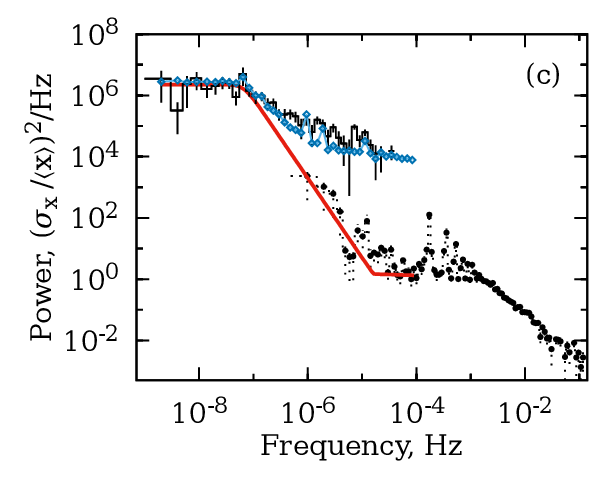}
	\caption{Modeling the influence of gaps and periodic variability on the shape of the power spectra for the monitoring observations. On each panel, the black step line indicates the observed power spectrum, the thick red (dark-gray) solid line represents the initial model, and the blue (light-gray) line represents the Monte Carlo model (similarly to Fig.~\ref{fig:asca_pds}). For comparison, the dots and dotted line indicate the ASCA power spectrum. On panel (a), the modeling was performed for the MAXI data; the initial model in which the flat portion extends to the lowest frequencies was used. On panels (b) and (c), a power-law power spectrum with an index of 2.6 was chosen as the initial model; the modeling for the MAXI (b) and ASM (c) data is shown. Here, we introduce the following notation (for comparison with Fig.~\ref{fig:maxi_mlfunc}): $\nu_{br}\sim 10^{-7}$~Hz is the position of the break in the flat spectrum at the lowest frequencies and $A \sim 10^5-10^6$ is the power level of the flat spectrum.}
	\label{fig:pds_long_sym}
\end{figure*}

The orbital peak has a noticeable width in both MAXI and RXTE/PCA power spectra. This broadening probably stems from the fact that the positions of the minima of the orbital cycle depend on the precession phase. It is well known \citep{PanfFabrRakh1997,Goranskij1998} that these minima are shifted, depending on the precession phase, and they become asymmetric. This effect is explained by the fact that the orientation of the accretion disk with respect to the donor star changes, and, as a consequence, the stellar heating conditions change.

Figure~\ref{fig:ph_prec} shows the precessional phase curves of \SSobj\ constructed from the MAXI and ASM observations using the ephemerides from \cite{Goranskij2011}. As with the power spectra, the amplitude of the ASM phase curve turned out to be higher than that of the MAXI one, which may be due to background overestimation in the ASM data. As expected, the maximum of the ASM phase curve occurs at phase $\psi=0$, but the MAXI phase curve turned out to be shifted by 0.1. Using the older ephemerides from \cite{Gies2002} and \cite{Goranskij1998} led to a similar result. According to the data by S.\,A.\,Trushkin (private communication), who continuously monitored \SSobj\ in the radio  band at the RATAN-600 telescope from early 2011 to late 2015, the activity of \SSobj\ occurs at the beginning of the precession maximum in eight of the eleven precession cycles over this period. Comparison of the radio light curve with the Swift/BAT X-ray data over the same period (Trushkin, private communication) shows that all eight precession maxima are asymmetric and are shifted to a later time. Since the MAXI observatory has been operating since 2009, we may conclude that the delay of the precession maxima from the MAXI data in Fig.~\ref{fig:ph_prec} is associated with the flare activity of \SSobj.

Thus, all three power spectra shown in Fig.~\ref{fig:pds_long_data} exhibit a gentle slope and a very large amplitude. They overlap with the ASCA power spectra in the frequency range $10^{-6} - 10^{-4}$~Hz. However, we find a discrepancy between them in amplitude by 2--4 orders of magnitude in this range. This difference in amplitudes may be related to the power leakage in the monitoring data due to the gaps in the observations. In addition, precession is the most powerful source of variability in the system (the highest peak in Figs.~\ref{fig:pds_long_data}a and~\ref{fig:pds_long_data}b); the X-ray flux from \SSobj\ changes by a factor of 2 over the precession cycle. To check whether this flattening of the power spectra is associated with the power leakage from the precession cycle, we performed modeling similar to what we did above for the ASCA data.

In our modeling we took into account the contribution of three periodic components to the observed variability: precession, nutation, and eclipses. We obtained the model of the precession trend by fitting the MAXI phase curve with two Fourier harmonics (at the fundamental frequency and twice as high, Fig.~\ref{fig:ph_prec}). We modeled the eclipses by a meander, a set of identical rectangular profiles with a constant period and a depth of 50\%. The dependence of the eclipse profile on the precession phase was disregarded. We took the orbital period to be 13.08223 days \citep{Goranskij2011} and the eclipse duration to be three days. We described the nutational variability by a sine wave \citep{Cherepashchuk2013} with the ephemerides $2443009.720271 + 6.287599\times E$ \citep{Davydov2008} and an amplitude of 10\%. The final model of the periodic trends was obtained by multiplying these three components. Otherwise, the algorithm for constructing the synthetic power
spectra was identical to that applied to the ASCA data. The modeling results are shown in Fig.~\ref{fig:pds_long_sym}.

In Fig.~\ref{fig:pds_long_sym}a (for the MAXI data), as the initial model we chose a power-law power spectrum with an index of 0.06 and a level of about unity in fractional rms squared normalization. This initial model corresponds to the case as if the flat portion observed in the ASCA data in Fig.~\ref{fig:asca_pds}b extend to $10^{-9}$~Hz. It can be seen from the figure that the synthetic power spectrum taking into account the periodic trends and gaps in the light curve is slightly lower than the observed one but considerably higher than the initial model. We obtained a similar result by applying the same initial model to the ASM and PCA data (not shown in the figure);  in the case of PCA, the agreement of the synthetic power spectrum with the observed one turned out to be even better. This suggests that the huge power observed in \SSobj\ at low frequencies is caused mainly by its periodic variability.

When analyzing the ASCA observations, the model in which the flat portion ended near $10^{-5}$~Hz and the power spectrum rose with an index of 2.6 at lower frequencies (Fig.~\ref{fig:asca_pds}a) turned out to be the best. We also applied the model rising with an index of 2.6 in the frequency range $10^{-7}-10^{-5}$~Hz (with a break at $\nu_{br}\sim 10^{-7}$~Hz and a flat spectrum with $A \sim 10^5 - 10^6$ below this break) to the monitoring data. The results are shown in Fig.~\ref{fig:pds_long_sym}b for MAXI and Fig.~\ref{fig:pds_long_sym}c for ASM.

\begin{figure}
	\center
	\includegraphics[width=0.5\textwidth]{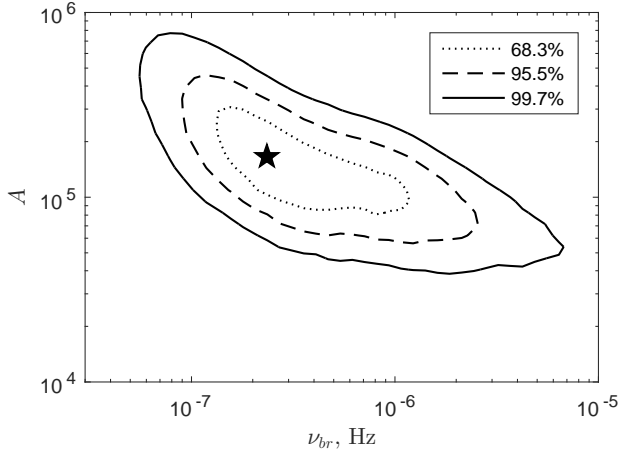}
	\caption{The contours corresponding to the 68.3\%, 95.5\%, and 99.7\% confidence intervals of the parameters: $\nu_{br}$ is the position of the break in the initial model of the MAXI power spectrum in Fig.~\ref{fig:pds_long_sym}b and $A$ is the power level of the flat spectrum. The star marks the values of the parameters that gave the best agreement of the model with the observations: $\nu_{br} \approx 2.4 \times 10^{-7}$~Hz and $A \approx 1.6 \times 10^5$ (in fractional rms squared normalization).}
	\label{fig:maxi_mlfunc}
\end{figure}

At frequencies above $10^{-7}$~Hz, the slope of all synthetic power spectra in Fig.~\ref{fig:pds_long_sym} turned out to be the same irrespective of the parameters of the initial model and roughly coincided with the observed one. This suggests that the power-law index of about unity (Fig.~\ref{fig:pds_long_data}) observed at these frequencies $\gtrsim10^{-7}$~Hz in the MAXI and ASM data is apparently associated with the large number of gaps leading to the power leakage from the precessional, orbital, and nutational peaks and does not reflect the real shape of the power spectrum. However, at lower frequencies ($10^{-7}$~Hz or lower), the monitoring data turned out to be quite sensitive to the level A of the initial model. Figure~\ref{fig:maxi_mlfunc} shows the contours of the parameters $A$ and $\nu_{br}$ for the MAXI data. $A=(1.6^{+0.9}_{-0.6}) \times 10^5$ in fractional rms squared normalization (the $1\sigma$ contour in Fig.~\ref{fig:maxi_mlfunc}) is required to properly describe the MAXI observations, with $nu_{br} \approx 2.4 \times 10^{-7}$~Hz. 

Thus, it follows from our simultaneous analysis of the MAXI and ASM monitoring data with the ASCA observations that the power spectrum of \SSobj\ in the frequency range $10^{-7}-10^{-5}$~Hz must rise sharply from the level about unity (the ASCA flat portion) to $\sim 10^5$. This rise is associated not only with the noticeable contribution from the power of the periodic variability included in the model but also with additional effects. Given that the time for the passage of material in the disk of \SSobj\ cannot be longer than half the nutation period, it is highly likely that this rise is due to changes in the accretion rate of material supplied by the donor.

\vspace*{1cm}
\section*{DISCUSSION}

\noindent Using RXTE/ASM and EXOSAT/ME data, \citep{Revn2006} found the power spectrum of \SSobj\ in a wide frequency range, from $10^{-8}$ to $10^{-2}$~Hz, to be described, on average, by a single power law with an index $\beta\approx = 1.5$. Our more detailed  analysis of the X-ray observations shows that the power spectrum of \SSobj\ has a more complex shape. It has a flat portion that is directly observed in Fig.~\ref{fig:asca_pds} and extends at least from $3 \times 10^{-5}$ to $10^{-3}$~Hz. The MAXI and RXTE/ASM monitoring observations at frequencies above $10^{-7}$~Hz show a gentle slope $\beta\approx 1$, which is mostly attributable to the periodic, mainly precessional,  variability of \SSobj. Because of the dominant contribution of the periodic component, we unable to confidently establish the shape of the power spectrum of the stochastic variability in this frequency range. However, a significant power is reliably observed in the monitoring data at frequencies below $10^{-7}$~Hz: higher than the level of the flat portion in the ASCA data by 5--6 orders of magnitude. This indirectly suggests that in the range $10^{-7}-10^{-5}$~Hz the flat portion ends and the power spectrum again has a power-law shape. 

\cite{Lyubarskii1997} showed that fluctuations of viscosity $\alpha$ could be responsible for the power-law power spectrum in a standard accretion disk. It is assumed that at various disk radii, these fluctuations occur independently and on time scales longer than the viscous time
\begin{equation}
\label{eq:time_visc}
t_{visc}(R)=\left[\alpha\left(H(R)\over R\right)^2 \Omega_K \right]^{-1},
\end{equation}
where $\Omega_K$ is the Keplerian velocity and $H$ is the disk half-thickness, have the pattern of white noise, i.\,e., have a flat spectrum. Viscosity fluctuations in each disk ring lead to a perturbation of the accretion rate in this ring. As the material passes through the disk, these perturbations accumulate; as a result, the power spectra of the accretion rate (the luminosity) turn out to be power-law.

We think that the same mechanism can be responsible for the stochastic variability of \SSobj. The flat portion presumably results from a sharp decrease in the viscous time in the supercritical disk. Previously \citep{AtapinSS433var2015}, we assumed that such an abrupt decrease occurs at the spherization radius, the place where the wind outflow begins, due to the increase in disk thickness. In the supercritical region, the disk is geometrically thin, $H\approxeq R$ \citep{ShakSun1973,Lipunova1999}; as a result, the viscous time becomes longer that the free-fall time only by  several times.

A more detailed analysis shows that the flat portion can be formed in the radiative disk zone, i.\,e., even before the spherization radius. In the radiative zone, where the radiation pressure dominates over the gas one, the disk thickness $H$ is constant \citep{ShakSun1973}. As a result, $t_{visc}\propto R^{7/2}$ in this zone. Such a strong dependence of the viscous time on the radius must also lead to a flattening of the power spectrum. The extent of the radiative zone of \SSobj\ is such that the viscous time in it drops from ten days at the outer boundary to $\sim 30$~s at the inner one.

On the other hand, the accretion disk of \SSobj\ can differ significantly from the standard one. The clear relationship between the nutational oscillations of the jets and the disk suggests that the time for the passage of material through the disk is shorter than the nutation half-period, i.\,e., less than three days \citep{Fabrika2004,Revn2006}. Such a rapid advance of material, hundreds of times faster than that in the standard disk, can be associated with spiral shock waves \citep{Sawada1986,Bisikalo1999}, efficiently taking the angular momentum away from the material. If the accretion disk is actually completely replenished every three days, then the shape of the power spectrum at frequencies below $\sim 4 \times 10^{-6}$~Hz must no longer be associated with the disk structure. The variability observed at these frequencies is then probably attributable to other processes, for example, a change in the amount of gas supplied by the donor star.

Yet another mechanism that can be responsible for the formation of a flat portion is the variability of the cooling jets. The jet luminosity provides the most of the observed X-ray flux from \SSobj\ \citep{Brinkmann2005,MedvFabr2010}). We showed previously that the jet model describes well the position of the high-frequency (at $10^{-3}$~Hz) break in the power spectrum and the slope of the power-law portion above the break \citep{AtapinSS433var2015}. We assumed the jets to consist of discrete clouds cooling either radiatively or adiabatically, depending on which mechanism is more efficient for a given initial density. The clouds were also assumed to be ejected randomly and to obey the Poisson distribution.

The jet variability model we proposed is basically the shot noise model. The shape of the power spectrum above the high-frequency break is determined by the profile of the time dependence of cloud temperature, while the position of the break is determined by the characteristic cooling time \citep{AtapinSS433var2015}. On time scales longer than the characteristic jet cooling time, the details of the cooling process cease to play a role, and the shape of the power spectrum is determined exclusively by the statistics of the ejection of clouds in the jet. Since we assumed that the clouds were ejected randomly and independently and that their average number ejected per unit time $<\dot{N}>$ was constant, the variability at all frequencies below the break had the pattern of white (uncorrelated) noise, i.\,e., a flat portion appeared in the power spectrum.

Calculations show that the jets in the supercritical disk are formed and collimated in its innermost parts, only several Schwarzschild radii from the black hole \citep{Ohsuga2011,Jiang2014}. The accretion rate in the supercritical disk region decreases linearly with radius $\dot{M}(r)=(r/r_{sp})\dot{M}_0$ due to the removal of material by the wind, while the spherization radius is proportional to the accretion rate \citep{ShakSun1973,Poutanen2007}. Therefore, the accretion rate variations at the outer disk boundary must lead only to a change in the spherization radius. In this case, approximately the same amount of material will reach the jet formation region irrespective of the initial accretion rate $\dot{M}_0$. The constancy of $<\dot{N}>$ may give rise to a flat portion, which is determined by the supercritical accretion
regime.

At the precession phases when the disk is seen edge-on, there is no flat portion in the power spectrum. At these phases, the X-ray flux is considerably weakened, and a single power-law spectrum with an index $\beta \approx 1.34$ is observed. Our modeling showed \cite{AtapinSS433var2015} that we observe the funnel radiation scattered by external clouds at these phases. At the precession phases when the disk is maximally turned toward the observer, the X-ray emission from \SSobj\ is mainly determined by the jets; therefore, we observe a flat portion in the power spectrum.

\section*{CONCLUSIONS}
\noindent

In this paper, we studied the X-ray variability of \SSobj\ at frequencies lower than $\sim 10^{-4}$~Hz. Using the uniquely long twelve-day ASCA observation allowed the power spectrum to be constructed in the frequency range from $10^{-6}$ to 0.1~Hz. To extend it to $10^{-9}$~Hz, we also analyzed the MAXI and RXTE/ASM data and the RXTE/PCA data.

Analysis of the ASCA observations confirmed the presence of a high-frequency break and a flat portion in the power spectrum of \SSobj\ at the precession phases when the funnel is maximally turned toward the observer. The derived parameters, the position of this break $\nu_2 = (1.54\pm 0.32) \times 10^{-3}$ and the slope of the power-law portion above the break $\beta_2 + \beta_3 = 1.68 \pm 0.09$, coincided, within the error limits, with those measured previously from RXTE data, suggesting a high stability of the observed picture.

When studying the low-frequency part of the power spectrum, we performed modeling that took into account the contribution to the observed variability from the eclipse and precession trend as well as the distortions due to the gaps in the light curve. The following parameters were obtained in the best fit model: the position of the low-frequency break $\nu_1 \approx 9.5 \pm 10^{-6}$~Hz and the slope of the portion below the break $\beta_2+\beta_1 \simeq 2.6$. It emerged that the variability associated with the eclipse and precession at low frequencies dominates over the stochastic variability, and this does not allow the errors of the model parameters to be determined. 

Similar modeling for the power spectra constructed from the monitoring data showed their shape to be determined to a considerable extent by the periodic variability of \SSobj. The power-law index at frequencies above $10^{-7}$~Hz turned out to be about unity. It is caused by power redistribution due to the large fraction of gaps in the observations and the low data quality. Nevertheless, a rise in power to $10^5-10^6$ with respect to the level of the flat portion in the ASCA data is reliably detected at lower frequencies.

It follows from our simultaneous analysis of the monitoring and ASCA data that in the frequency range from $10^{-7}$ to $10^{-5}$~Hz the flat portion must end, and the power spectrum must rise. However, the jet nutation which is directly related to a force moment acting on the outer edges of the disk by the donor star, suggests that the time for the material passage is no more than three days. This characteristic time roughly corresponds to the position of the low frequency break $\nu_1 \sim 10^{-5}$~Hz. We cannot firmly establish the frequency of this break, because the significant increase in power at frequencies below $\nu_1$ is associated mainly with the periodic processes in the system. For this reason, we conclude that the variability at frequencies below $10^{-5}$~Hz is probably associated not with the accretion disk structure but is attributable, apart from the periodic processes, to other factors, for example, a change in the amount of gas supplied by the donor. This process can also be periodic; it is associated with the nodding clocks of \SSobj. The gravitational field of the donor perturbs the outer parts of the accretion disk every 1.5 days: in the disk plane and across the disk. This process is quite complex, but, nevertheless, it is clear that the accretion rate in the disk will change noticeably. The X-ray flux from \SSobj\ is completely determined by the disk inclination to the line of sight, which also complicates the variability of this object at frequencies below $\sim 10^{-5}$~Hz.

Viscosity fluctuations in the accretion disk can be responsible for the stochastic variability of \SSobj. The flat portion in the power spectrum appears due to a sharp decrease in the viscous time in the supercritical and radiative disk zones. The X-ray flux from \SSobj\ in the energy range 1--7~keV is completely determined by the cooling jets, which consist of separate gas clouds. Irrespective of the accretion rate onto the outer disk edge, the mass flux in the X-ray jets is approximately constant, because the supercritical disk controls the internal accretion rate, getting rid of the excess gas in the form of a wind. Our additional argument is that at a constant outflow rate in the jets of \SSobj\ the clouds can be ejected randomly according to shot noise statistics. The external conditions during the jet formation, a sharp decrease in the time for the passage of material through the disk in the supercritical and radiative zones, can determine this process.

\vspace*{0cm}
\section*{ACKNOWLEDGMENTS}
We are grateful to the anonymous referees for their valuable remarks that allowed the text of the paper to be improved. This research is based on the MAXI data provided by RIKEN, JAXA and the MAXI team as well as on quick-look results provided by the ASM/RXTE team. The work was supported by the Russian Science Foundation (project no. 14-50-00043, analysis of monitoring observations), the Russian Foundation for Basic Research (project no.~16-02-00567 and 16-32-00210), was financed in part by the Program for Support of Leading Scientific Schools of Russia (NSh-7728.2016.2). S.\,N.\,Fabrika thanks the Program of Competitive Growth of the Kazan Federal University.
 

\setlength{\bibsep}{0pt}

\end{document}